# Finite size effects on surface excess quantities and application to crystal growth and surface melting of epitaxial layers

**P. Müller**
*Centre de Recherche sur les Mécanismes de la Croissance Cristalline[1]*
*Campus de Luminy, case 913, F-13288 Marseille Cedex 9, France*

**Abstract**

From a macroscopic viewpoint phase transitions as surface melting or growth mode can be described in terms of Gibbs excess quantity duly amended by size effect. The aim of this study is to consider such amended quantities to describe surface melting and Stranski-Krastanov transition of epitaxial layers. The size effect so introduced allow to predict the equilibrium thickness of the wetting layer of Stranski Krastanov growth mode and to describe and classify two different melting cases: the incomplete premelting relayed by a first order transition and the continuous premelting relayed by continuous overheating.

***Keywords:*** *Surface energy, adhesion energy, size effects, Stranski-Krastanov, Surface melting*

## Introduction:

From a classical thermodynamic point of view phase transitions as surface melting of semi-infinite crystals or crystal growth can be partially described in terms of Gibbs excess quantity duly amended by size effects (since usual Gibbs excess quantities are only well defined for semi-infinite systems).The aim of this study is to generalise this approach in order to describe two dimensional (2D) towards three dimensional (3D) transition (Stranski-Krastanov or SK transition) of epitaxial layers as well as the surface melting of epitaxial layers.

For this purpose in a first section we define surface excess quantities for finite size slabs useful to describe the SK transition (section I1). Then we define (section I.2) the surface and interfacial quantities for the more complex case of composite slabs useful for the description of surface melting of epitaxial layers and give some experimental evidences (section I.3).

Concerning Stranski Krastanov transition (section II) we define a thermodynamical model to describe the conditions necessary for the transition (section II1) and show how in near equilibrium conditions a good description of the size effect allow to calculate the thickness of the wetting layer and the activation energy of the transition. Nevertheless the smaller the lattice mismatch between the deposit and its substrate, the greater the activation energy so that for couples A/B with small lattice mismatch another transition mechanism has to be described (section II2).

---
[1]Associé aux Universités Aix-Marseille II et III.

Concerning surface melting (section III) we will show how the use of finite size quantities allow us to describe and classify two different melting cases: the incomplete premelting relayed by a first order transition and the continuous premelting relayed by continuous overheating in agreement with more complex numerical calculations.

## I/ Size effect on surface excess quantities

### I.1/ Simple slab ($n_i$ layers)

As first described by Gibbs [1] all bulk extensive properties can present some excess at the surface. Nevertheless all these excess quantities, as surface and adhesion energies $\gamma_i^\infty$ and $\beta_i^\infty$ of a solid i, are well defined only for semi-infinite solids. For a finite solid i that only contains a few layers $n_i$, surface and adhesion energies depend on the number of layers of the solid. Indeed, considering not only short range "chemical interactions" but also longer ranged ones as $-r^{-6}$ dispersion forces or $-e^{-r/\zeta a}/r$ for screened Coulomb forces, to the chemical bonding between the layers of a slab of thickness d there adds $d^{-3}$ or $e^{-d/\zeta a}$ contributions respectively. For screened Coulomb forces the surface energy $\gamma_i(n_i)$ of a finite slab having $n_i$ layers thus reads

$$\gamma_i(n_i) = K \sum_{n=0}^{n_i-1} e^{-n/\zeta_i} = K \frac{1-e^{-n_i/\zeta_i}}{1-e^{-1/\zeta_i}} = \gamma_i^\infty \left(1-e^{-n_i/\zeta_i}\right) \qquad (1)$$

Where the material constant K characterising the chemical interactions has been asymptotically linked to the usual surface energy $\gamma_i^\infty$ of semi-infinite solid by writing $\gamma_i(n_i \to \infty) = K/1-e^{-1/\zeta_i} = \gamma_i^\infty$. The same reasoning can be made for the adhesion energy $\beta(n_i,n_j)$ of a film A (thickness $n_A$) on a substrate B (thickness $n_B$) which is connected to the interfacial bonding [2]. In this case the summation has to be made on the layers of the film A then on the layers of the substrate B so that $\beta(n_A,n_B)$ reads [2]

$$\beta(n_A, n_B) = \beta^\infty \left(1-e^{-n_A/\zeta}\right)\left(1-e^{-n_B/\zeta}\right) \qquad (2)$$

where $\beta^\infty$ is the usual adhesion energy in between the two semi-infinite solids (for simplicity we do not distinguish $\zeta$ for A, B or AB long range interactions). The interfacial energy $\gamma_{AB}(n_A,n_B)$ can now be obtained from the extended Dupré relation [3,4] $\gamma_{AB}(n_A,n_B) = \gamma_A(n_A) + \gamma_B(n_B) - \beta(n_A,n_B)$ and thus reads:

$$\gamma_{AB}(n_A,n_B) = \gamma_A^\infty\left(1-e^{-n_A/\zeta}\right) + \gamma_B^\infty\left(1-e^{-n_B/\zeta}\right) - \beta^\infty\left(1-e^{-n_A/\zeta}\right)\left(1-e^{-n_B/\zeta}\right) \quad (3)$$

**I2/ Composite slab ($n_i/n_j/n_k$)**

The adhesion energy $\beta_{i/jk}$ of a material i ($n_i$ layers) over a *composite slab* constituted by $n_j$ layers of material j over $n_k$ layers of material k can be obtained by means of a thermodynamic process where the 3-composite slab $i(n_i)/j(n_j)k(n_k)$ is obtained as a combination of single 2-composite slabs as: $i(n_i)/j(n_j)k(n_k) = i(n_i)/j(n_j) + i(n_i+n_j)/k(n_k) - i(n_j)/k(n_k)$. Thus there is [4]

$$\beta_{i/jk}(n_i,n_j,n_k) = \beta_{i/j}^\infty\left(1-e^{-n_i/\zeta_i}\right)\left(1-e^{-n_j/\zeta_j}\right) + \beta_{i/k}^\infty e^{-n_j/\zeta_j}\left(1-e^{-n_i/\zeta_i}\right)\left(1-e^{-n_k/\zeta_k}\right) \quad (4)$$

We can thus define the interfacial energy of material i ($n_i$ layers) onto the composite material $\gamma_{i/jk}(n_i,n_j,n_k)$ by using again Dupré's relation, so that

$$\gamma_{i/jk}(n_i,n_j,n_k) = \gamma_i(n_i) + \gamma_j(n_j) - \beta_{i/jk}(n_i,n_j,n_k) \quad (5)$$

Obviously equations (2) and (3) can be recovered from equations (4) and (5) with $n_i=n_A$, $n_j=n_B$ and $n_k=0$.

Lastly lay stress on the fact that owing to Shuttleworth' relation [5] all these size corrections also apply to surface and interfacial stress defined as excess quantities [6].

**I3/ Experimental evidences:**

In our opinion such size effects have been clearly put in evidence by the experimental study of the asymptotic behaviour of stress establishment in thin films [7,2]. Indeed though usually the in-plane force exerted by a 2D film of material A epitaxially supported by a lattice-mismatched thick film of material B is written $\sum(h_A) = \sigma h_A + \Delta s$ (where σ is the average stress in the film A of thickness $h_A$ and $\Delta s$ a contribution caused by the excess of stress at surfaces and interfaces), the true in-plane force must be written $\sum(h_A) = \sigma h_A + \Delta s^\infty\left(1-e^{-h_A/\zeta a}\right)$ where $\Delta s^\infty$ is the surface and interfacial contribution for semi infinite slab A and $\left(1-e^{-h_A/\zeta a}\right)$ the size correction factor of relation (1) where a is an atomic unit. Thus in the case of perfectly pseudomorphous film where the bulk contribution $\sigma h_A$ can be easily calculated (provided the epitaxial misfit and the elastic constants are known), the surface contribution $\Delta s^\infty\left(1-e^{-h_A/\zeta a}\right)$ can be easily extracted form experimental measurements of $\sum(h_A)$ recorded by the cantilever method. It is the case of Ge/si(001) experiments [7] where the surface stress contribution has been extracted and shown to be well fitted by $\Delta s^\infty \left(1 - e^{-h_A/\zeta a}\right)$ with a$\zeta$=0.28 nm and $\Delta s^\infty$=2.3 Jm$^{-2}$ [2] (see figure 1).

In the following we will use such amended excess quantities in order to revisit some well known epitaxial growth and surface melting phenomena.

## II/ Application to crystal growth: the Stranski Krastanov transition

Let us recall that three possible mechanisms of epitaxial growth have been recognised [8]: the three dimensional or Volmer-Weber growth, the layer by layer or Frank van der Merwe growth and the layer by layer growth followed by three dimensional growth or Stranski-Krastanov (SK) growth.

In absence of misfit and near equilibrium conditions Bauer [8] rationalised these growth modes by defining the wetting energy

$$\Phi_\infty = 2\gamma_A^\infty - \beta^\infty \qquad (6)$$

It was then shown [9] that three dimensional (3D) growth *near equilibrium* conditions only takes place for $\Phi_\infty > 0$ and at supersaturation $\Delta\mu > 0$ (where $\Delta\mu$ is the chemical potential difference per atom between the deposited crystal A and a reservoir of A) whereas two dimensional (2D) growth only takes place for $\Phi_\infty < 0$ and at undersaturation $\Delta\mu < 0$. In absence of misfit 2D and 3D growth modes thus are well differentiated but the condition for SK growth is not clear. It is not the case when the lattice mismatch between the deposited crystal A and its foreign substrate B is duly considered. In this case we will show that in absence of surface stress, the near equilibrium growth conditions are: $\Phi_\infty > 0$ with $\Delta\mu > \varepsilon_0^{3D}$ for 3D growth and : $\Phi_\infty < 0$ with $\Delta\mu < \varepsilon_0^{2D}$ for 2D growth where $\varepsilon_o^{2D}$ and $\varepsilon_o^{3D}$ are the elastic energy density per atom in two dimensional layer and three dimensional crystal respectively. Thus since, owing to the elastic relaxation of the lateral faces a 3D crystal $\varepsilon_o^{3D} < \varepsilon_o^{2D}$, it can be believed that in near equilibrium conditions SK growth could take place at $\Phi_\infty \approx 0$ in a reduced domain of supersaturation $\varepsilon_o^{3D} < \Delta\mu < \varepsilon_o^{2D}$.

*In this section we want to precise the conditions of SK growth and discuss the problem of the activation energy that has to be overpass for SK transition.* For this purpose we will define a thermodynamic process by which a 3D crystal may grow on 2D layers, seek for equilibrium conditions and discuss activation energy involved in SK transition.

**II.1/ Free energy change for SK transition**

The deposit A is considered to be obtained from the condensation of a perfect vapour onto a lattice mismatched semi-infinite crystal B. (A and B are cubic with parameter a and b and supposed not to mix). The epitaxy is with parallel axis on a (001) plane and the (001) surfaces and interface of A and B are supposed to be stable. The in-plane misfit being defined by $m_o = (b-a)/a$. The final state of the condensation is a 3D crystal of volume V sitting on z pseudomorphic layers over the substrate B (Stranski Krastanov case) (see figure 2). Furthermore we will only consider box shaped 3D crystals and we will neglect surface stress.

The free energy change of the SK condensation is composed of three terms: **(i)** The chemical work to form (on an area $L^2$) a 2D film of z layers and an island (volume $V = h\ell^2$). It reads $\Delta F_1 = -\Delta\mu(zaL^2 + h\ell^2)$ where $\Delta\mu$ is the supersaturation per unit volume of vapour A and a an atomic linear size. **(ii)** The work of formation of the surfaces of the crystal A followed by its adhesion on the bare substrate B: $\Delta F_2 = \Phi_\infty\left[(L^2 - \ell^2)(1-\exp(-z)) + \ell^2\right] + 4\gamma'_A h\ell$ where $\Phi(z) = 2\gamma_A(z) - \beta_{AB}(z) = \Phi_\infty(1-\exp(-z/\zeta))$ is the size dependent wetting factor and where we have neglected $\exp(-(z+h/a\zeta))$ against $\exp(-z/\zeta)$ [10]. In the following we will take $\zeta = 1$. **(iii)** The total elastic energy stored by the composite system $\Delta F_3 = \varepsilon_o\left[zaL^2 + h\ell^2 R(h,\ell,z)\right]$ (with $\varepsilon_o = Ym_o^2$ and Y a combination of elastic constants) which is the sum of the homogeneous energy stored by z pseudomorphous layers (thickness a, lateral size L) and the elastic energy of the 3D upperlying crystal of volume $V = h\ell^2$. The factor $0 < R(h,\ell,z) < 1$ is a relaxation factor describing the elastic relaxation of the 3D upperlying crystal. It has to be calculated for each specific case (see for example [11]). Finally the total energy change $\Delta F = \Delta F_1 + \Delta F_2 + \Delta F_3$ can be written [10]:

$$\Delta F = -\Delta\mu(V + L^2 za) + \Phi_\infty\left[\left(L^2 - \left(\frac{V}{r}\right)^{2/3}\right)(1-e^{-z}) + \left(\frac{V}{r}\right)^{2/3}\right] + 4\gamma'_A V^{2/3} r^{1/3} + \varepsilon_o(VR + zaL^2) \quad (7)$$

where we introduce the aspect ratio $r = h/\ell$ of the 3D crystal and write $R(h,\ell,z) = R$.

Notice that if the 2D layers have to be formed, A must wet B so that $\Phi_\infty$ must be negative.

The equilibrium state is found by minimisation of the total energy change ΔF. The zeros of the partial derivatives of (7) in respect with z, V and r give the equilibrium values of z, V and r noted z*, V* and r* respectively. They are reported in table I

| | |
|---|---|
| $\left.\dfrac{\partial \Delta F}{\partial z}\right\|_{V,r}=0$ | $z^* = \ln\left[\dfrac{|\Phi_\infty|}{\left(\varepsilon_o\left(1+\dfrac{h}{a}\theta^2\dfrac{\partial R}{\partial z}\right)-\Delta\mu\right)a}(1-\theta)\right] \approx \ln\left[\dfrac{|\Phi_\infty|}{(\varepsilon_o-\Delta\mu)a}\right]$ |
| $\left.\dfrac{\partial \Delta F}{\partial V}\right\|_{z,r}=0$ | $V^*(r) = \left[\dfrac{\dfrac{8}{3}\gamma'_A - \dfrac{2}{3}\dfrac{|\Phi_\infty|}{r}e^{-z}}{\Delta\mu - \varepsilon_o R}\right]^3 r$ |
| $\left.\dfrac{\partial \Delta F}{\partial r}\right\|_{z,V}=0$ | $r^{*2/3} = -\dfrac{4\gamma'_A}{3\,\varepsilon_o}V^{-1/3}\left[1-\dfrac{r_0}{r}\exp(-z)\right]\left(\dfrac{dR}{dr}\right)^{-1}$ |

**Table I:** *equilibrium values $z^*,V^*,r^*$ where $\theta=(\ell/L)^2$ defines the 3D coverage. The approximated relation is valid for weak coverage*

Notice that the expressions of $z^*$ and $V^*$ show that there is an interplay in between the wetting layer and the 3D crystal. More precisely the greater the volume of the 3D crystal, the smaller the number of underlying layers what is supported by both experimental evidences [12,13] and numerical calculations as well [14]. Thus one can adopt another point of view [15] to describe the SK transition by considering that owing to the energy gain due to the elastic relaxation of the 3D crystals, some of the upper layers of a metastable 2D strained film of thickness $z'$ can spontaneously transform into stable 3D islands supported by $z<z'$ underlying layers. The free energy change $\Delta F'$ of this transition reads $\Delta F' = \Delta F(z,V) - \Delta F(z',0)$ that means for $\theta^2 h/a \ll 1$

$$\Delta F' = -\varepsilon_o V[1-R(r)] + |\Phi_\infty|\left[V - \left(\dfrac{V}{r}\right)^{2/3}\right]e^{-z/\zeta} + 4\gamma'_A V^{2/3} r^{1/3} \qquad (8)$$

## II.2/ Discussion

### II.2.1/ Layer by layer growth

Here we are only concerned with Frank-van der Merwe growth that means $\Phi_\infty < 0$ for having 2D condensation with $V=0$ and thus $\theta=0$ so that from table I there is $z^* = \ln\left[\dfrac{|\Phi_\infty|}{(\varepsilon_o-\Delta\mu)a}\right]$. Since z must be positive, the z layers can only exist for $-\infty < \Delta\mu < \varepsilon_o$. Thus *for having 2D growth, the supersaturation cannot overpass the bulk elastic energy density stored in the strained layers*[2]. In figure 2 we plot the free energy density $\Delta F/L^2$ as a function of z for different chemical potentials $\Delta\mu$. $\Delta F/L^2$ shows minima, at z=z*, for $\Delta\mu < \varepsilon_o$. In this case since $(\Delta F/L^2)_{z^*} < 0$, 2D layers form spontaneously provided $\Phi_\infty < 0$ that means each layer z is a 2D phase, built at a given undersaturation $\Delta\mu_z = \varepsilon_o - |\Phi_\infty/a|\exp(-z)$. Up to saturation $\Delta\mu = 0$ there builds up a finite number of layers $z_o = \ln(|\Phi_\infty|/a\varepsilon_o)$ that only depends on the wetting over strain energy ratio $|\Phi_\infty|/a\varepsilon_o = |2\gamma_A - \beta|/[Ym_o^2]$. This result is largely experimentally supported on very different pairs A/B: reversible multilayers adsorption measurements (for a review see [16])[3]. However let us note that 2D growth may also take place at $\Delta\mu > \varepsilon_o R$. Indeed for $\Delta\mu > \varepsilon_o R$ the two members $\Delta F$ (eq. (7)) are negative so that $\Delta F/L^2$ is always negative and is a decreasing function of the number of layers z. In these conditions each supplementary layer decreases $\Delta F$ but $\Delta F$ has no minimum so that the equilibrium layers number can no more be defined (see figure 3) so that we will call these conditions out of equilibrium conditions.

**II.2.2/ 2D relayed by 3D growth:**

Since for SK transition, z* and V* must be positive, from table I, Stranski Krastanov growth can only occur for:

$$\varepsilon_o R < \Delta\mu < \varepsilon_o \qquad (9)$$

that means in a finite domain of chemical potential $\Delta\mu$ (excepted when elastic relaxation does not play so that $R=1$). In figure 4 we schematically plot the number of 2D layers as a function of the chemical potential $\Delta\mu$. To each layer formation z* corresponds a constant value of $\Delta\mu$ given by $\Delta\mu(z^*) = \varepsilon_o - e^{-z^*}|\Phi_\infty/a|(1-\theta)$ but 3D islands may appear as soon as $\Delta\mu > \varepsilon_o R$. The smallest volume reached by a 3D upperlying crystal can be obtained by injecting $\Delta\mu = \varepsilon_o$ in the expression of V* in table I. For a given aspect ratio r and for $\Phi_\infty e^{-z} \to 0$ this minimum volume reads [10] : $V^*_{\min}(r) = [(8\gamma'_A/3)/\varepsilon_o(1-R)]^3 r$ which becomes infinite when elastic relaxation is not considered !

---

[2] In absence of misfit ($m_o=0$) the usual condition for 2D growth $\Delta\mu<0$ is recovered

[3] Let us note that for $\Delta\mu = \varepsilon_o$, z* becomes infinite but the elastic energy stored diverges so that the system has to relax either by plastic deformation or islanding [10].

The activation energy for Stranski-Krastanov transition $\Delta F'^*(r)$ can thus be obtained by injecting the equilibrium values V* and r* of the table I for $\Phi_\infty < 0$ in the equation (8). For $\Phi_\infty e^{-z} \to 0$ it reads:

$$\Delta F'^* = \frac{\left((8/3)\gamma'_A\right)^3}{(\Delta\mu - \varepsilon_o R)^3} r^* \left[\frac{3}{2}\Delta\mu - \varepsilon_o\left(1 + \frac{R}{2}\right)\right] \quad (10)$$

Since $\varepsilon_o R < \Delta\mu < \varepsilon_o$ the smaller value of the activation energy is reached for $\Delta\mu = \varepsilon_o$ that means for $\Delta F'^*(r^*, z^*) = \frac{4}{3}\frac{(4\gamma'_A)^3}{(3\varepsilon_o)^2}\frac{r^*}{[1-R]^2}$ [10]. As usually the activation barrier $\Delta F'^*(r)$ is proportional to $\gamma'^3_A$ but $\varepsilon_o^{-2}(1-R)^{-2} \propto m^{-4}(1-R)^{-2}$ obviously plays the role of a driving force. At the limit $z \to \infty$ R can be analytically calculated [11] so that for Cu(111) where $\gamma \approx 1300$ ergcm$^{-2}$ and $\varepsilon_o/m_o^2 = 2.3 \cdot 10^{-12}$ ergcm$^{-3}$ [17] the minimum value of $\Delta F^*$ reaches the values $\Delta F^*/kT \approx 100$ for $m_0=1\%$, $\Delta F^*/kT \approx 30$ for $m_0=2\%$ but $\Delta F^*/kT \approx 2$ for $m_0=8\%$) [10].

*Thus due to the upper limit of $\Delta\mu$ the activation energy for nucleating the supported 3D crystal only catches reasonable values (a few kT) for sufficiently high misfit ($|m|>2\%$).* However since some cases of SK growth are well known for smaller misfit, some authors argue of the too high activation energy to consider that the SK growth is a pure matter of kinetics [18-21], nevertheless electrochemical studies have clearly shown that SK growth can take place in near equilibrium conditions [22]. Some other authors try to find a mechanism by which the activation energy could be lowered. It could be classically the case of adsorption on the 3D facets which lowers γ', faceting of the instable 2D layers, or more exotic phenomena as sequential nucleation of islands and pits [23] somewhat similar to Asaro-Tiller-Grienfeld instability, and even the appearance of a liquid layer at the 2D/3D interface [24] !.

However in our opinion the activation energy of SK transition can be lowered if the 2D underlying layers do not grow in near equilibrium conditions. Indeed we have seen in section II.2.1 that 2D layers can grow even for $\Delta\mu > E_o$ condition. In this case the growth can start by 2D growth at $\Phi_\infty < 0$ so that each new 2D layer lowers the free energy change. One can no more define an equilibrium number of 2D layers ($z^* \to \infty$) but for high positive value of the supersaturation $\Delta\mu$ the equilibrium volume of the 3D islands can become very small (a few atoms) so that the activation energy for the transition becomes negligible. The number of underlying layers thus should be fixed by kinetics and evaluated by the following procedure when the accessible SK activation energy is fixed to a certain amount of kT. Indeed let us suppose that SK transition occurs when the activation energy $\Delta F^*=n\ kT$. It is then formally possible to plot the abacus $\Delta F^*(r^*,z)$ given by (10) as a function of z and thus to find the value of z for which $\Delta F^*=n\ kT$ and thus for which SK transition starts. It is thus very important to be able to calculate the z dependence of the relaxation factor R that means the Green function describing the effect of a point force on a composite slab (z layers of material A on a semi infinite material B). Nevertheless up to now and to the best of our knowledge such z dependence of relaxation factor R has not been calculated so that no predictions can be made for the moment. However we believe that our proposition of *non equilibrium* 2D growth at $\Delta\mu>E_o$ but $\Phi_\infty<0$ followed by 3D growth near equilibrium conditions is a quite simple solution to obtain SK transition for very weak values of misfit.

## III/ Application to surface melting

Substrate S of material B now bears a lattice-mismatched composite material A of $n_s$ solid layers and $n_l$ liquid layers (fig.). The $n_s$ layers are in pseudomorphous contact and epitaxially stressed by S whereas its $n_l$ upper layers A are in the liquid state.

Our model of surface melting uses the notion of finite size surface and interfacial specific energies as justified by [25].

### III.1/ Free energy of the system

Our purpose being to seek for the equilibrium number of liquid layers as a function of temperature[4], we have to minimise a thermodynamic potential of the composite system constituted of $n = n_l + n_s$ layers of A sitting on a semi-infinite substrate S (see fig.5). We will note this system l/s/S. Therefore we have to minimise the Gibbs free energy per solid mole :

$G= G=N_s G^s + N_l G^l + G^{surf}$ where $G^s$ is the Gibbs free energy per solid mole (number $N_s$ per unit area), $G^l$ the Gibbs energy per liquid mole (number $N_l$ per unit area). Taking into account the size effect via equations (4) and (5), the excess energy due to surface and interfaces $G^{surf}$ reads [4]:

$$G^{surf}(n_s, n_l) = N_{avo}\left[2\gamma_S + (2\gamma_l - \beta_{l/s})(1-e^{-n_l/\zeta_l}) + (2\gamma_s - \beta_{s/S})(1-e^{-n_s/\zeta_s}) + (\beta_{l/s} - \beta_{l/S})e^{-n_s/\zeta_s}(1-e^{-n_l/\zeta_l})\right]$$

---

[4] The melting point of S ($T_S$) is much higher that the melting point $T_m$ of A so that experiments can be done around $T_m$ without alteration of B and A.

**III.2/ Equilibrium conditions:**

The equilibrium number of liquid layers can be obtained by the condition $\partial G/\partial N_l|_N = 0$. For essentialness, we neglect the size differences of $v^S$ and $v^l$ and suppose $\zeta_s = \zeta_l = \zeta$, so that $\partial G/\partial N_l|_N = 0$ reads:

$$\Delta U(T) - T\Delta S(T) - \frac{V^s E m^2}{1-\nu} + N_{av}\frac{b^2}{\zeta}\left[\Phi e^{-n_l/\zeta} - \Gamma e^{-n_s/\zeta}\right] = 0 \qquad (11)$$

where $\Delta U = U^l - U^s$, $\Delta S = S^l - S^s$ are the melting energy and entropy respectively. The constants $\Phi$ and $\Gamma$ read:

$$\Phi = 2\gamma_l - \beta_{l/s} \equiv \gamma_l + \gamma_{ls} - \gamma_s \qquad (12)$$

$$\Gamma = (2\gamma_s - \beta_{s/S}) + (\beta_{l/S} - \beta_{l/s}) \equiv \gamma_{sS} + \gamma_{ls} - \gamma_{lS} \qquad (13)$$

where all specific energies are those of the semi-infinite phases with planar surfaces $\gamma_i$, $\gamma_{ij}$ or $\beta_{i/j}$ (we omit $\infty$ subscript). According to the sign of the factors $\Phi$ and $\Gamma$ there follows two new premelting cases we analyse more clearly in the following discussion.

Since at the bulk melting point $T_m$ (without stress) there is $\Delta U_m(T_m) = T_m \Delta S_m(T_m)$, neglecting the heat capacity change at constant pression $\Delta C_m$ one has not too far from the melting point the linear dependence $\Delta U(T) - T\Delta S(T) \approx \Delta S_m(T_m)(T_m - T)$ where $\Delta S_m(T_m)$ is the latent melting entropy at the melting point $T_m$. In the following we will note the melting latent entropy $\Delta S_m$. The equilibrium condition thus reads:

$$T_m' - T = -\frac{N_{av} b^2}{\Delta S_m \zeta}\left[\Phi e^{-n_l/\zeta} - \Gamma e^{-n_s/\zeta}\right] \qquad \text{with} \qquad T_m' = T_m - \frac{E m^2}{1-\nu}\frac{V^s}{\Delta S_m} \qquad (14)$$

where $T_m'$ defines the melting point of the strained film. Indeed since $\Delta S_m > 0$ *an epitaxial coherent strained film melts at a lower temperature than a strain free film*. This shift may be important since for typical values $\Delta S_m$ = 2 cal mole$^{-1}$deg$^{-1}$, E=10$^{11}$ ergcm$^{-3}$, $\nu$=1/3, V$^s$= 20 cm$^3$ mole$^{-1}$ the shift is 3.7, 15, 60 K for misfits of 1, 2 or 4 % respectively.

Finally for the easiness of the discussion let us consider the second derivative of G in respect to $N_l$ at constant T and N:

$$\partial^2 G/\partial N_l^2\Big|_{T,N} = -\frac{N_{av} b^2}{\zeta^2}\left[\Phi e^{-n_l/\zeta} + \Gamma e^{-n/\zeta} e^{n_l/\zeta}\right] \qquad (15)$$

**III.3. Discussion:**

The condition Φ>0 being precluded for surface melting [26], two cases can be encountered according to the sign of Γ. Indeed according to the definition (13) Γ<0 may be written $\gamma_l + \gamma_{ls} + \gamma_{sS} < \gamma_l + \gamma_{lS}$ what means the l/s/S system is preferred to the l/S one. Therefore if Φ<0 and Γ<0 the interface l/s is pushed away from both the liquid surface and the substrate S. These two forces thus may balance each other. In the case Φ<0 and Γ>0 the two effective forces act in the same sense so that there is no more balance and the l/s interface must be attracted to the substrate S. We will see that the case Φ<0 and Γ<0 leads to a continuous increase of the number of liquid layers with temperature whereas in the case Φ<0 and Γ>0 some instability from continuous to discontinuous behaviour occurs.

### III.3.1/ Φ<0 and Γ<0: the surface induced continuous premelting and superheating:

From (15) there is $\partial^2 G / \partial N_l^2 \big|_{T,N} > 0$ for all values of $n_l$, so that in all the domain $0<n_l<n$, there is continuous stable melting from the dry point $T_s$ (the temperature above which the first liquid layer forms) to $T_l$ (the temperature where the last solid layer melts) defined by

$$T_s = T_m' + \frac{N_{av}}{\Delta S_m} \frac{b^2}{\zeta} \left[ \Phi - \Gamma e^{-n/\zeta} \right] < T_m' \quad \text{and} \quad T_l = T_m' + \frac{N_{av}}{\Delta S_m} \frac{b^2}{\zeta} \left[ \Phi e^{-n/\zeta} - \Gamma \right] > T_m' \quad (16)$$

So that for a thick enough film $T_s$ depends only upon the wetting factor Φ whereas $T_l$ only depends only upon Γ. *The domain of continuous melting thus has to be called premelting when $T<T_m'$ or overheating when $T>T_m'$.* This continuous melting at astride $T_m'$ can be explicitly calculated by solving the quadratic equation:

$$\Phi e^{-2n_l/\zeta} + \frac{\Delta S_m \zeta}{N_{av} b^2} (T_m' - T) e^{-n_l/\zeta} - \Gamma e^{-n/\zeta} = 0 \quad (17)$$

The result is given on figure 6 (the used data are given in the figure caption)

### III.3.2. Φ<0 and Γ>0: the surface induced two stage premelting:

In this case, it can be seen from (15) that $\partial^2 G / \partial N_l^2 \big|_{T,N}$ may have positive or negative values according to the value of $n_l^* = \frac{n}{2} - \frac{\zeta}{2} \ln\left(\frac{\Gamma}{|\Phi|}\right)$ where $dn_l/dT\big|_{n_l^*} = \infty$.

Let us describe the surface melting as a function of temperature that means for increasing number of liquid layers from $0<n_l<n$.

* For $0<n_l<n_l^*$, equation (17) with now $\Gamma>0$ has two solutions. A stable one $n_l^{stable}<n_l^*$ (corresponding to a minimum of G) where $\partial^2 G/\partial N_l^2\big|_{T,N}>0$ and an unstable one $n_l^{unst.}\geq n_l^*$ (corresponding to a maximum of G) where now $\partial^2 G/\partial N_l^2\big|_{T,N}\leq 0$. These two solutions continuously meet at the temperature $T^*=T_m'-\dfrac{N_{av}b^2}{\Delta S_m \zeta}2\sqrt{|\Phi\Gamma|}e^{-n/2\zeta}$ [4]. Since at this point $\partial^2 G/\partial N_l^2\big|_{T,N}=0$ any further increase of temperature produces an irreversible first order melting at $T^*<T_m'$.

* For $n_l>n_l^*$, there is only a stable solution corresponding to a continuous premelting from $T_s$ to $T_l$ (see figure 5)

Therefore when $\Phi<0$, $\Gamma>0$ there is *a two stage premelting*. The first stage roughly concerns half the film ($0<n_l<n_l^*$) which continuously melt. The second stage ($n_l>n_l^*$) corresponds to a first order melting at $T^*<T_m'$. (see figure ). Thus in a first stage the equilibrium thickness of the premelted liquid continuously increases with temperature, then below some critical temperature $T_c<T_m$ the solid the slab melts completely (first order transition). Let us note that such two stage premelting has been recently predicted by numerically calculations [27]. Obviously a more complete discussion for specific systems needs thermodynamic data on surface and interfacial quantities. (For a more complete discussion see [4])

**Conclusion**

Though Landau's theory of phase transitions can be expressed in very general terms of order parameter, relations of quantitative interest can only be obtained by adjoining models. In this more classical approach of some phase transitions we use some macroscopic quantities (measurable from semi-infinite phases) but duly amended from size effect by means of an interlayer potential. It is thus possible to describe and discuss the physics of some phase transitions of finite size systems. The so-obtained results are in quite good agreement with experiments as well as other more complex numerical approaches but clearly give access to the physical meaning in a quite simple form.

**Acknowledgements:** Pr R.Kern is acknowledged with gratitude for very fruitful discussions

**Figure captions:**

**Figure 1:** Stress change versus thickness h (in angstroms) recorded during Ge/Si(001) growth [7]. Diamonds: experimental results $\Sigma(h_A)$; straight line: bulk contribution $\sigma_A h_A$; dots: surface contribution $\Sigma(h_A) - \sigma_A h_A$; continuous line: fit of surface contribution by $\Delta s^\infty \left(1 - e^{-h_A/\zeta a}\right)$ [2].

**Figure 2:** Stranski Krastanov schema (3D crystal on 2D wetting layer supported by a substrate B) before (left) and after (right) elastic relaxation of the 3D upperlying crystal.

**Figure 3 :** **a/** Free energy change $\Delta F/L^2$ for layer growth versus the number of wetting layers z for different chemical potentials $\Delta\mu$. Beyond $\Delta\mu = \varepsilon_o$ the curves no more exhibit a minimum excepted for $z^* \to \infty$. **b/** Number of equilibrium layers versus the chemical potential $\Delta\mu$. Beyond $\Delta\mu = \varepsilon_o$, z* tends towards infinity.

**Figure 4:** Number of equilibrium layers z* versus the chemical potential Dm in case of SK growth. For $m_o \neq 0$, 3D islands may appear as soon as $\Delta\mu > \varepsilon_o R < \varepsilon_o$.

**Figure 5:** Schematic drawing of surface induced melting of epitaxial layers. Substrate S of material B, deposited film (thickness n) of material A with $n_s$ solid layers and $n_l = n - n_s$ liquid layers.

**Figure 6:** Premelting $n_l$ versus T. Four cases with the same wetting parameter $\Phi < 0$. **(1)** Usal premelting of a thick solid ($n_s = \infty$) reaching asymptotically $T_m'$ the bulk melting point of the strained material. **(2)** premelting of thin solid film (n layers): $\Gamma = 0$: same premelting curve as **(1)** but ending in $n_l = n$; case $\Gamma < 0$ n-finite, melting astride $T_m'$ with its overheating zone $T > T_m'$; $\Gamma > 0$ n-finite, premelting going over continuously at $T^* < T_m'$, followed by first order premelting at $n_l \approx n/2$. (Calculations have been performed for: n=8, $\Phi$=-50 erg cm$^{-2}$, $\Gamma$=-50, 0, +50 erg cm$^{-2}$.)

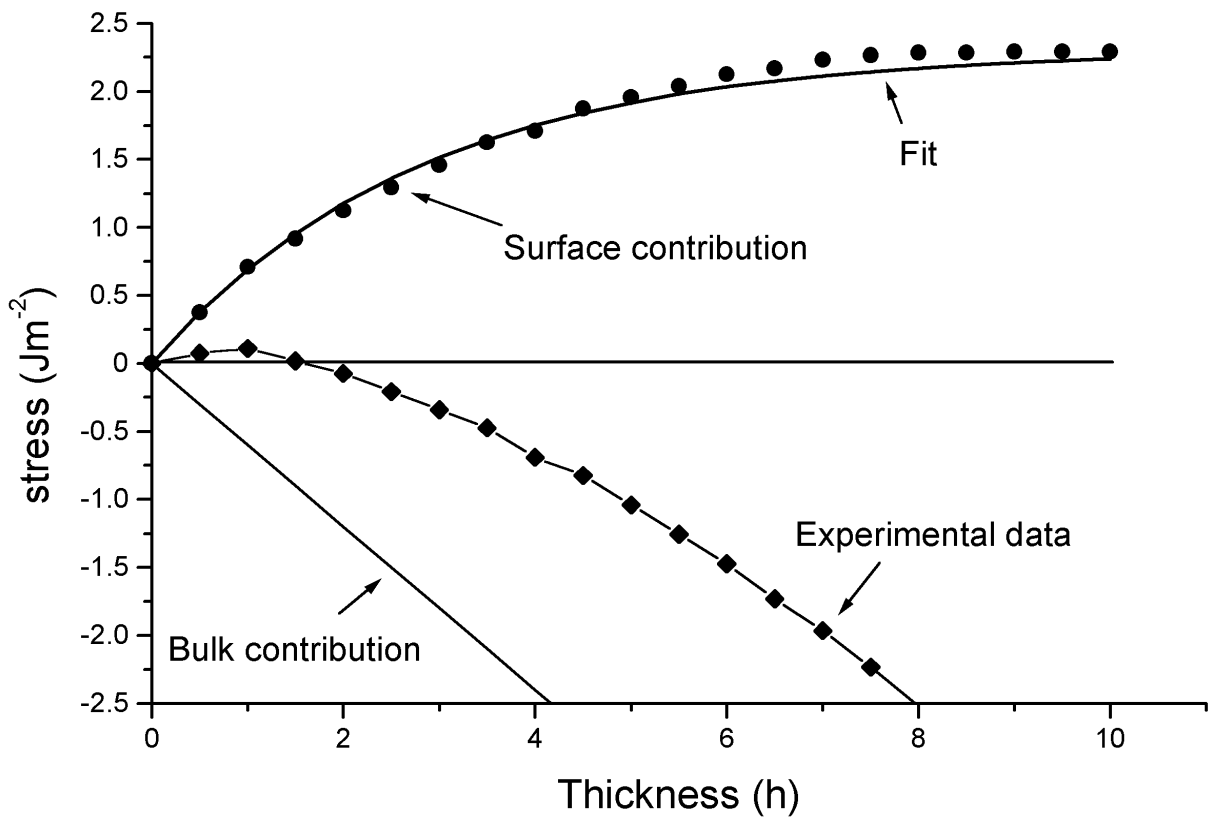

**Figure 1:**

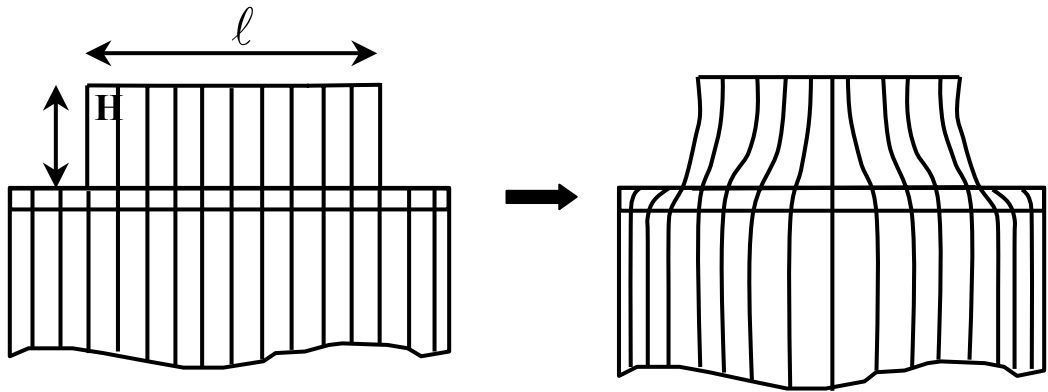

**Figure 2:**

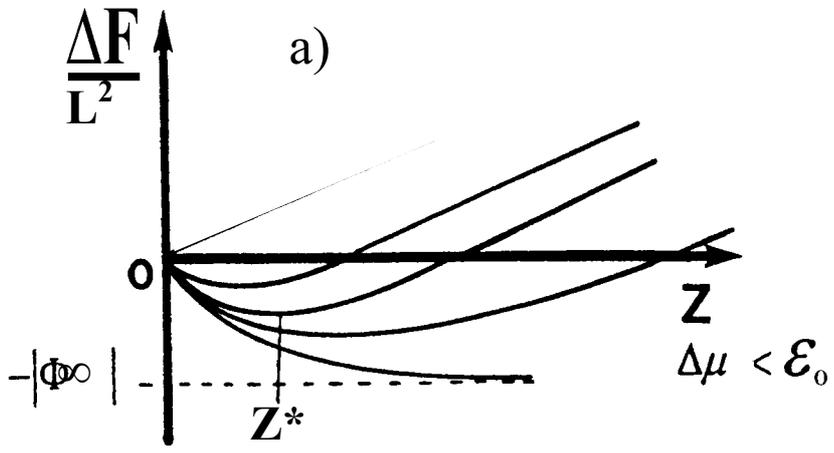

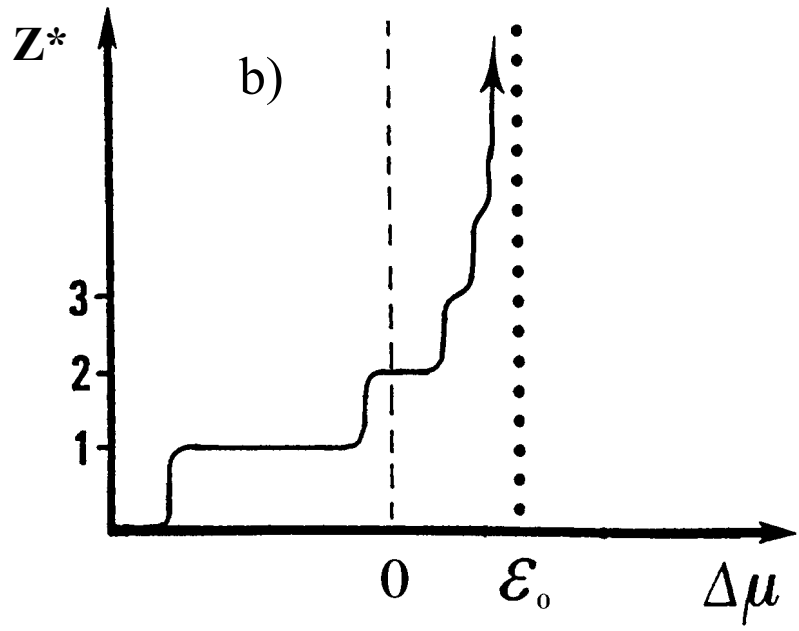

**Figure 3**

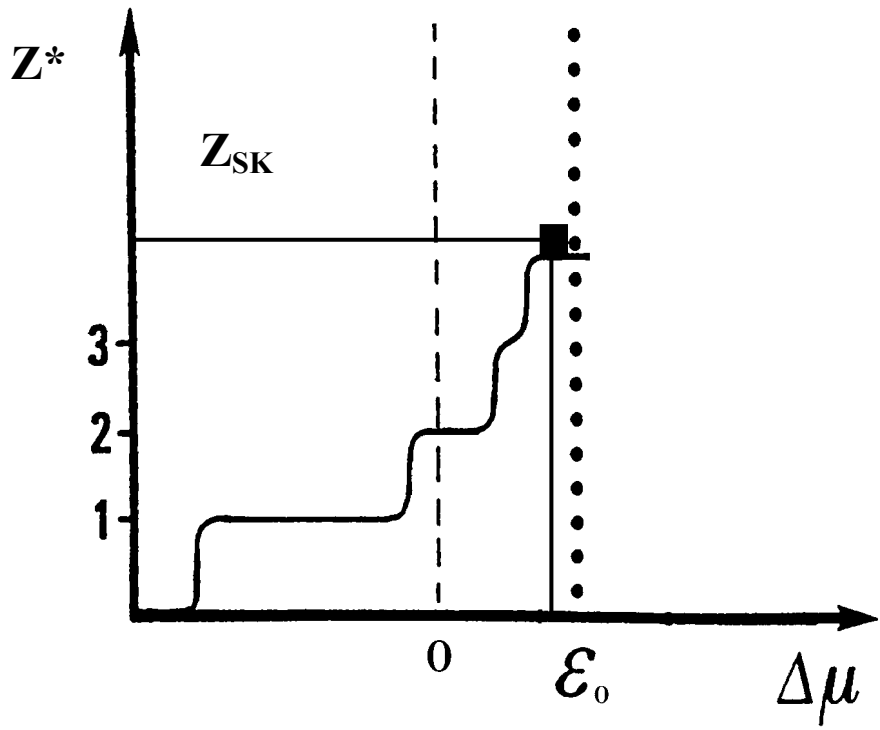

**Figure 4:**

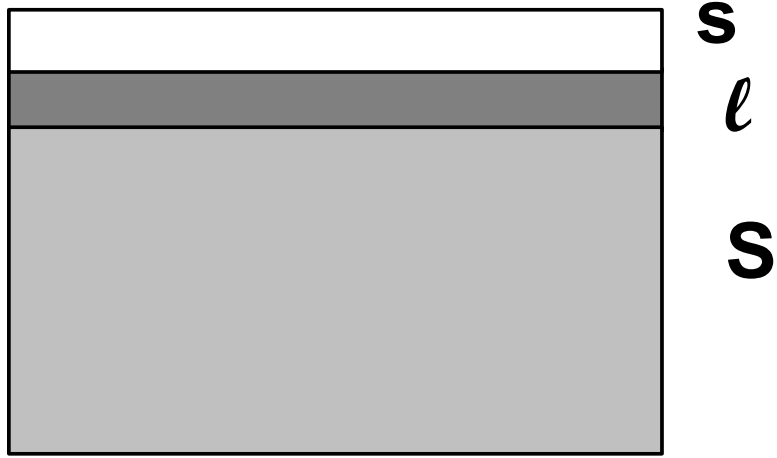

**Figure 5:**

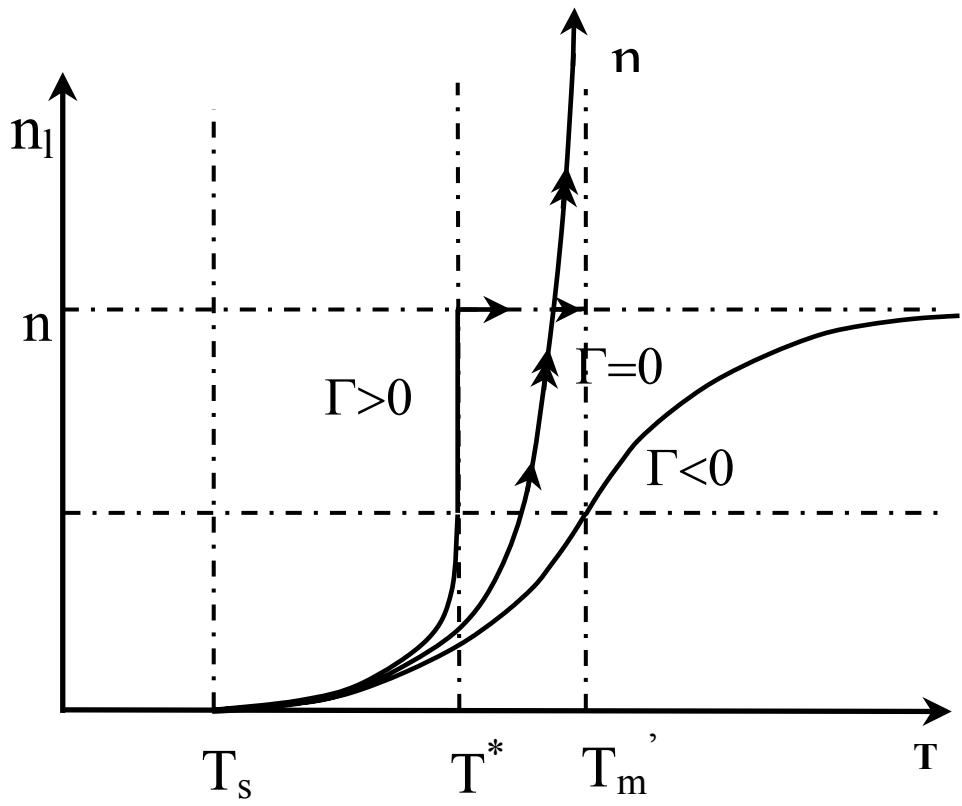

**Figure 6:**